\documentclass{kluwer}  
\usepackage{epsfig}
\newdisplay{guess}{Conjecture}

\begin{document}                                                                                   
\begin{article}
\begin{opening}         
\title{Morphology and Evolution of Galactic and Magellanic Cloud
Planetary Nebulae} 
\author{Letizia \surname{Stanghellini}}  
\runningauthor{Letizia Stanghellini}
\runningtitle{Planetary Nebula Morphology and Evolution}
\institute{
Space Telescope Science Institute, 3700 San Martin Drive,
Baltimore MD 21218.\newline
Affiliated to the Astrophysics Division, Space Science
Department of ESA.
}
\begin{abstract}
Planetary nebulae (PNe) exist in a range of different morphologies, 
from very simple and symmetric round shells, to elliptical, 
bipolar, and even quadrupolar shapes. They present extremely complex
ensembles of filaments, knots, ansae, and shell multiplicity. 
It is then 
overwhelmingly complicated to derive reasonable evolutionary paths to justify
the observed shapes of PNe.
The confrontation between the evolution of the
shells and that of the central stars is needed to understand the 
origin of the morphological variety. 
We present some background and recent results on the correlations between 
PN morphology and PN nuclei (PNNi) evolution, 
including a study on the 
Magellanic Cloud PNe.
\end{abstract}
\end{opening}  
         
\section{Introduction}

Planetary nebulae originate from stellar ejecta during the 
Thermal Pulse phase on
the Asymptotic Giant Branch (TP-AGB). Stars that do not reach the
Chandrasekhar core mass 
evolve through the AGB phase and beyond, thus setting a stringent limit
to the PN progenitor mass, at least if no mass transfer occurs.
After the envelope ejection via slow ($\approx 20 ~{\rm km ~s}^{-1}$) and very 
massive ($10^{-5}$ to $10^{-4} ~{\rm M_{\odot}~ yr}^{-1}$) 
superwind, the star evolves
toward higher temperatures at constant luminosity, substained by 
stellar wind and
nuclear burning in the hydrogen (or, in some cases, helium) shell, 
and eventually the stellar UV photons
ionize the ejecta. After the nuclear burning quenches, the star evolves at nearly 
constant radius and fades to white dwarf luminosities. 

This evolutionary scheme is widely accepted. Nonetheless, there are several
nearly unexplored aspects of post-AGB evolution that may create 
confusion when comparing data and theory. First, the exact path of
stellar evolution after the onset of the TP-AGB phase is not totally
clear. To date, still questionable is the mass-loss treatment
used in evolutionary calculations (e.g., Vassiliadis \& Wood 1994, Bloecker 1995); 
in debate is the 
mass-luminosity relation, which steepens for very
massive progenitors, due to hot-bottom burning (Bloecker \& Schoenberner 1991), as well as for lower mass stars evolving with overshooting
(Hervig {\it et al.} 1998).
Second, the transition time lag between the superwind quenching and the PN illumination
can not be tracked in evolutionary calculation, since the mass-loss
law is not known. Third, the occurrence of hydrogen- versus helium-burning
PNNi is not completely understood, although the onset of helium
burning in post-AGB stars has been extensively studied (Iben {\it et al.} 1983). 

The field of AGB and post-AGB modeling is in full bloom, and new evolutionary
calculation, together with synthetic evolutionary models (e.g., 
Marigo {\it et al.} 1999, Stanghellini \& Renzini 1999 {\it in preparation}) 
are contributing to build 
a consistent theoretical scenario. On the other hand, hydrodynamical
models, pioneered by Franz Kahn (1983), have been updated to
compare to observed PNe. The stellar and shell models, in the end, have
to reproduce the evolution of stars and nebulae as complex systems, 
thus PNe and central stars should be studied together.

Planetary nebula morphology carries information on the shell ejection,
the central star energetics, and the surrounding medium.
Asymmetry in PNe must be ascribed to asymmetries in the formation
mechanism, as well as to their evolution. 
For example, a companion star or planet 
has proven effective to build up material in the equatorial plane of the 
system, thus enforcing a bipolar outflow when the envelope
ejection occurs. Other types of morphological substructures
such as fliers and knots have been observed in many PNs. Such structures 
can be explained in detail by dynamical evolution (e.g., Dyson, this
volume). Multiple shell PNe have been interpreted as 
multiple ejection or dynamically evolved PNe 
(Stanghellini \& Pasquali 1995).
Other phenomena like post-AGB stellar pulsation or magnetic field
can be also responsible for other asymmetries.

 \section{Correlations between PN morphology, stellar 
astrophysics, nebular evolution, and stellar populations.}

In this paper we will limit the analysis to main PN morphologies:
round (R), ellipticals (E), bipolar (B), pointsymmetric (P), and 
quadrupolar (Q) (see Manchado {\it et al.} 1996, MGSS). 
The first correlations that use Galactic PN 
morphology as an {\it independent variable} have been derived a few decades ago,
when Greig (1972),
found that binebulous PNe were closer to the Galactic plane than
other morphological types;
Peimbert and
collaborators (Peimbert 1978, Peimbert and Torres-Peimbert 1983, Calvet 
\& Peimbert 1983) found a trend between bipolar shape and enhanced 
helium and N/O abundances; Zuckermann \& Aller (1986)
found that the enrichment of CNO elements
anti-correlates with the  PNe altitude on the Galactic plane. 

In 1992 the first large morphological catalog of Galactic PNe was published
(Schwarz {\it et al.} 1992, SCM). Several studies on morphology versus other
parameters, based on the catalog images, 
followed (e.g. Stanghellini {\it et al.} 1993). Most of these studies, together with the
older analysis based on smaller samples, are consistent with the 
following scenario: round PNe originate from the lower mass stars, 
located randomly within the Galaxy; asymmetric (especially bipolar) 
PNe have massive progenitors, they are chemically enriched to show
dredged-up elements, and they have low Galactic altitude. Elliptical PNe
seem to have intermediate properties. Asymmetric PNe have been found to
be optically 
thicker than symmetric PNe. 

The above results, if inspiring, have nonetheless two 
biases. First, the SCM catalog is not complete and homogeneous, it
does not include all PNe within the Observatory range. 
Second, most of the correlations described 
are based on stellar properties derived
by assuming that the statistical distances to Galactic PNe (Cahn {\it et al.} 1992)
are reliable. To alleviate the first bias, a new set of observations has 
been recently undertaken (MGSS), providing
a homogeneous and complete database. Several 
studies are under way based on this catalog, a preview of whose is 
presented in $\S$2.1. 

Circumventing the distance bias is difficult, 
since only a dozen or so Galactic PNe have measured 
reddening, or cluster membership,
distances. 
A way to check some of the earlier results without the distance bias
is to study 
Magellanic Cloud PNs (MCPNe), whose morphology can be seen almost exclusively with 
the use of HST. In $\S$2.2 we show some of the MCPNe morphological correlations.

\subsection{Galactic PNe}

We introduce in this paper
a sample of the correlations found with the MGSS 
morphological database.
In most cases we group B and Q PNe together, given their
morphological similarities.
Figure 1 shows the correlations between H I 
and He II Zanstra PNN temperatures. 
The closer the two temperatures, the thicker is the PN to the ionizing UV flux.
We see that, on average, more B (and Q) 
PNe are closer to the 1:1 line, thus they are 
optically thicker than R and E PNe. 
Nonetheless, the correlation is weaker than in Stanghellini 
{\it et al.} (1993). We conclude that the result is still uncertain.

From the location of central stars on the HR diagram, we were able
to make a rough estimate of the PNNi masses, whose histograms are plotted
in Figure 2. The histogram bins are built accordingly to the 
evolutionary tracks by Vassiliadis and Wood
(1994; M/M$_{\odot}$=.55, .57, .68, .9).
This Figure confirms that most of the high mass PNNi 
are hosted by asymmetric PNe.

\begin{figure}
\centerline{\epsfig{file=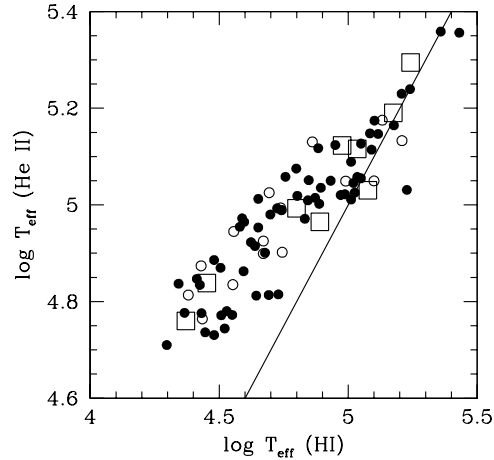,width=20pc}}
\caption{Optical thickness of round (open circles), elliptical (filled
circles), and
bipolar/quadrupolar (open squares) PNe: comparison of H and He II 
Zanstra temperatures. The solid line represents the 1:1 relation.}
\end{figure}

\begin{figure}
\centerline{\epsfig{file=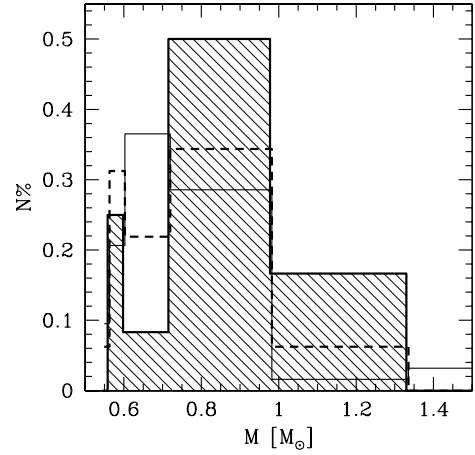,width=20pc}}
\caption{Central star's mass distribution of round (dashed line), 
elliptical (thin line), and
bipolar/quadrupolar (shaded histogram) PNe.}
\end{figure}

\begin{figure}
\centerline{\epsfig{file=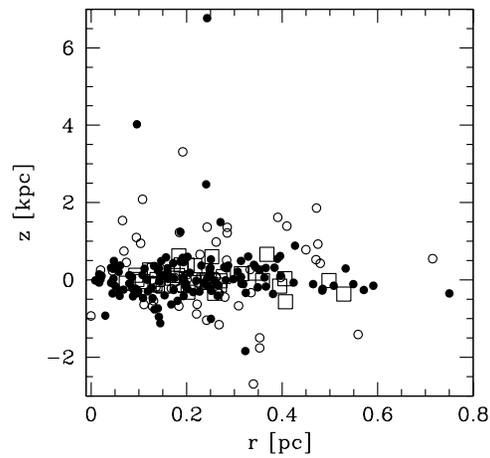,width=20pc}}
\caption{Distance of PNe from the Galactic plane, versus nebular radius.
Symbols as in Fig. 2.}
\end{figure}

In Figure 3 we examine the Galactic distribution of PNe according to their
altitude on the Galactic plane, plotted against nebular
radius. The result is that bipolar
PNe are located, on average,  at lower altitudes than other morphological classes, 
hinting to a different stellar population.

The results of Figures 1 through 3
are in agreement with the older finding that bipolar PNe 
have massive progenitors and belong to a different stellar population 
than elliptical or round PNe. Before confirming this scenario, 
we must also
examine whether there is a fundamental bias in the observation
of bipolar versus other PNe. By analyzing the extinction to different types
of PNe, we find that bipolar and quadrupolar PNe are heavily extincted, 
while round PNe have, in general, lower extinction. If the observed
extinction of all morphological types were mostly external, 
then we are sampling different volumes in the 
Galaxy for each morphological type. In order to solve this question, we
plan to study the internal to external extinction ratios 
for a sample of bipolar and round/elliptical PNe in the Milky Way.
\subsection{Magellanic Cloud PNe}

MCPNe can be  
resolved at the optical wavelengths 
via space astronomy, and offer a way to study  
PNe and their evolution in a distance-bias free environment.
HST images of MCPNe have been acquired and studied by 
Blades {\it et al.} (1992), Dopita {\it et al.} (1996),
Vassiliadis {\it et al.} (1998), and Stanghellini {\it et al.} (1999, SBOBL). 
To date, only 27 MCPNe have published resolved images from space,
thus we lack a statistically significant sample to extend and explore the 
results obtained for Galactic PNe. Nonetheless, it has been noted that (1)
MCPN morphological classes are the same as in Galactic PNe; (2) 
all bipolar PNe are type I (high N/O, an indication of third dredge-up
occurrence) and all round PNe are non-type I; (3) the fading time
of round PNe is longer than for other morphologies, indicating 
lower mass progenitors. 

\begin{figure}
\centerline{\epsfig{file=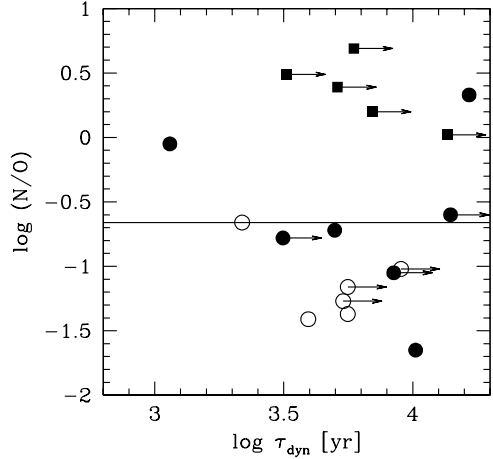,width=20pc}}
\caption{Third dredge-up enrichment in MCPNe. 
Round PNe: open circles; Elliptical
PNe: filled circles; bipolar PNe: filled squares. The line divides type I
(top half of the plot) and non-type I PNe 
(from SBOBL).}
\end{figure}

\begin{figure}
\centerline{\epsfig{file=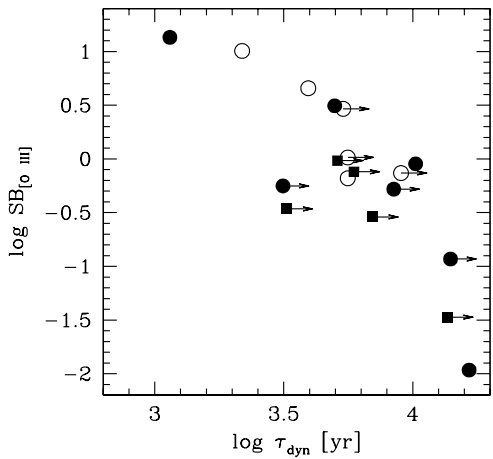,width=20pc}}
\caption{Magellanic Cloud PNe with known morphology: the evolution of
the [O III] surface brightness. Symbols as in Fig. 4 (from SBOBL).}
\end{figure}

The correlation of chemical enrichment with morphology is shown in Figure 4.
The observed N/O ratio is 
plotted against the dynamical time of the PNe for each main morphological
group. We find a striking confirmation of the segregation of bipolar 
PN as type I. In addition we find 
that all round PNe of this sample are non-type I. Elliptical PNe belong to
both groups. We do not detect any evolutionary enhancement of N/O. 
This result is an important link between progenitor mass, chemical enrichment,
and morphology of PNe. It states that since all bipolar PNe have enhanced N/O,
they have gone through the third dredge-up during the AGB. It is classically
assumed that the third dredge-up is related to the massive stars, 
although recent work implies that the third dredge-up
phenomenon may involve a larger mass range (Hervig et 
al. 1998). 

In Figure 5 we show the decline of the [O III] surface brightness with 
dynamical time for MCPNe of various morphological types. This Figure 
suggests that round PNe of this sample have low nuclear mass. 
In fact, the post-AGB evolution of the surface brightness depend on the 
central star fading time, which is directly correlated with the mass 
of the star.

\section{Summary and Future Projects}

Recent analysis of Galactic and Magellanic Cloud PNe have generally
confirmed that PN morphology is 
a tracer of stellar population and progenitor mass. On the other hand, the
new, homogeneous Galactic database (MGSS)
has posed the problem of space distribution
of the different PN morphologies, which need to be solved before we can 
ultimately confirm our findings.

The spatially-resolved MCPNe, on the other hand, 
are still too few to have a sound confirmation of the Galactic
results.   
We are planning an extended study of PNe in the Large Magellanic Cloud
via (Cycle 8) STIS/HST slitless spectroscopy. We will obtain 
stellar and nebular high-resolution information on a sample of at least 50 
PNe, 
with the goal of attach statistical significance to the correlations
described in this paper.

\acknowledgements

Thank to Eva Villaver for her help in the data compilation. Thanks to 
Richard Shaw and to the SBOBL collaborators for discussions 
on Magellanic Cloud PNe.

\end{article}

\begin{thebibliography} 


\bibitem{}Blades J. C., {\it et al.} 1992, ApJ 398, L44

\bibitem{}Bloecker, T. 1995, A\&A 299, 755

\bibitem{}Bloecker, T., \& Schoenberner , D. 1991, A\&A 244, L43

\bibitem{}Cahn, J. H., Kaler, J. B., \& Stanghellini, L. 1992, A\&AS 
94, 399

\bibitem{}Calvet N., \& Peimbert M. 1983, Rev. Mex. Astron. Astrof. 5, 319

\bibitem{}Dopita, M. A., {\it et al.} 1996, ApJ 460, 320

\bibitem{}Greig W. E. 1972, A\&A 18, 70

\bibitem{}Hervig, F., Schoenberner, D., \& Bloecker, T. 1998, A\&A 340, L43 

\bibitem{}Iben I. Jr, Kaler J. B., Truran J. W., \& Renzini A. 1983, ApJ 264, 
605

\bibitem{}Kahn F. D. 1983, in ``Planetary Nebulae'', ed. D. R. Flower (Dordrecht: Reidel), 305

\bibitem{}Manchado, A., Guerrero, M., Stanghellini, L., \& Serra--Ricart, M. 
1996, ``The IAC Morphological Catalog of Northern Galactic Planetary Nebulae'' 
(Tenerife: IAC) (MGSS)

\bibitem{}Marigo, P., Girardi, L., \& Bressan, A. 1999, A\&A 344, 123

\bibitem{}Peimbert M. 1978, in ``Planetary Nebulae: Observations and Theory'', 
ed. Y. Terzian (Dordrecht: Reidel), 215

\bibitem{}Peimbert M., \& Torres--Peimbert S. 1983, in ``Planetary Nebulae'', 
ed. D. R. Flower (Dordrecht: Reidel), 233

\bibitem{}Schwarz H. E., Corradi R. L. M., \& Melnik J. 1992, A\&AS 96, 23 (SCM)

\bibitem{}Stanghellini L., Corradi R. L. M., \& Schwarz H. E. 1993, 
A\&A 279, 521

\bibitem{}Stanghellini, L., \& Pasquali, A. 1995, ApJ 452, 286

\bibitem{}Stanghellini, L., Blades, C. J., Osmer, S. J., Barlow, M. J., \&
Liu, X.-W. 1999, ApJ 510, 687 (SBOBL)

\bibitem{}Vassiliadis E., {\it et al.} 1998, ApJ 503, 253

\bibitem{}Vassiliadis, E, \& Wood, P. R. 1994, ApJS 92, 125

\bibitem{}Zuckerman B., \& Aller L. H. 1986, ApJ 301, 772

\end{thebibliography}
\end{document}